\documentclass
[12pt]
{article}
\usepackage{latexsym,amsfonts}

\makeatletter
\@addtoreset{equation}{section}
\makeatother

\def\be{\begin{equation}}
\def\ee{\end{equation}}
\def\bea{\begin{eqnarray}}
\def\eea{\end{eqnarray}}
\def\nn{\nonumber}

\def\a{\alpha}
\def\b{\beta}

\def\de{\delta}
\def\e{\epsilon}           

\def\f{\phi}               

\def\h{\eta}

\def\l{\lambda}

\def\p{\pi}                

\def\x{\xi}

\def\L{\Lambda}

\begin{document}

\title{{\bf Non-commutative Euclidean structures in compact spaces}}

\author{
B.-D. D\"orfel\\
Institut f\"ur Physik, Humboldt-Universit\"at zu Berlin\\
Invalidenstra\ss e 110, D-10115 Berlin, Germany\\}

\date{January 14, 2000}

\maketitle

\begin{abstract}
  Based on results for real deformation parameter $q$ we introduce a compact
  non-commutative structure covariant under the quantum group $SO_q(3)$ for $q$
  being a root of unity. To match the algebra of the $q$-deformed operators
  with necesarry conjugation properties it is helpful to define a module over
  the algebra generated by the powers of $q$. In a representation where $X^2$
  is diagonal we show how $P^2$ can be calculated.  To manifest some typical
  properties an example of a  one-dimensional $q$-deformed Heisenberg algebra
  is also considered and  compared with non-compact case.  
  \vskip 0.5cm
  

  \noindent
  Keyword(s): non-commutative geometry, quantum group, root of unity
\end{abstract}
\section{Introduction}
In paper [1] it was shown, how the $q$-deformation of the well-known group
$SO(3)$ to quantum group $SO_q(3)$ can be used to define a non-commutative
quantum space as a comodule of the quantum group. It is very natural to exploit
the $ R$ matrix as the main tool. Its decomposition into projectors generates
a non-commutative (three-dimensional) Euclidean space of coordinates.

In all papers known to us the non-commutative structure has been defined for
real $q$ only. The value of $q$ becomes important when we demand hermiticity
for coordinates (and later on for momenta). For general complex $q$ the $R$
matrix looses its hermiticity which requires a new definition of conjugation
for the coordinate operators. On the other hand their are at least two reasons
why one should investigate the case of complex $q$ .
First, real $q$ implies always a non-compact coordinate space, while for a
compact space we have to admit complex values of $q$ . In context with the
fact, that non-commutative geometry [2] is considered to be the result of some
deep dynamical principle which may be found e.g. in string theory the case
of compactified dimensions is of special interest. We start here the
consideration of an example with only compactified coordinates. The more
interesting case with compact and non-compact dimensions (which seems to
require different $q$) is due to further work. Second, we know the quantum
group $SO_q(3)$ for generic $q$ and especially the case $q$ being a root of
unity, where it demonstrates some pecularities [3,4]. It is therefore
interesting how a non-commutative quantum space can be constructed in that
special case. This will be the main aim of our paper.

As we have already mentioned, the key point is the definition  of a
conjugation for coordinates and momenta, which are later required to be
self-adjoint with respect to that conjugation. Different conjugations result
in different spaces and hence different physics. The conjugation we will
propose below is of course equivalent to ordinary conjugation for real $q$
. We know two ways which are both consistent with $SO_q(3)$ . The choice that
fits best with our problem is the one, where $q$ is left untouched during
conjugation. Thus if $\bar{X}$ is the conjugate of an operator $X$ , the
conjugate of   $qX$ is $q\bar{X}$ . This choice has been used already before,
e.g. in [3]. To do this in a mathematical correct way we define a right module
over the algebra generated by all powers of $q$ with the additional condition
for some power to equal $-1$ (see next Chapter). The other way , one may find i.e. in [4], seems to work better in
case if one deals with non-hermitean operators having only real eigenvalues,
which will not be the case here.

At the first moment our definition looks rather unnatural but in Chap. 2 we
shall describe how it works and mention the consequences. The most important
one of them is that self-adjoint operators will have (instead of real ones)
eigenvalues which are real functions of the parameter $q$ . But this is just
what we need, because the scaling operator and its commutation properties
force coordinates and momenta to have eigenvalues proportional to powers of
$q$ .

The paper is organized as follows. In Chap. 2 we recall the basic formulae for
the quantum space of $SO_q(3)$ and state the modifications for our $q$ . In
Chap. 3 we consider a one-dimensional example of a $q$-deformed Heisenberg
algebra and demonstrate how it works for $q$ being a root of unity. It is
rather helpful to compare our results with earlier ones for real $q$ with the
same example. In our main Chap. 4 the non-commutative space covariant under
$SO_q(3)$ is considered and matrix elements of coordinates and momenta are
calculated. The results are presented explicitly and do not contain any
divergencies which usually occur if one simply replaces $q$ in formulae
derived earlier for real $q$ only. 
 
\section{Euclidean phase space for §q§ being a root of unity}
First we have to recall some basic formulae of the non-commutative space from
paper [1] which do not depend on the nature of $q$ .
The $R$ matrix of $SO_q(3)$ is decomposed like 
\be \hat R = P_5 - {1 \over {q^4}}P_3+ {1 \over {q^6}}P_1 
\ee
We shall not give the projectors $P_i$ here, because we need only $P_3$ . 
The non-commutative Euclidean space is defined by:
\be
P_3XX=0
\ee
In the common basis (2.2) looks like :
\bea X^3X^+ &  = &  q^2X^+X^3 \nn   \\ 
X^3X^- &  = &  q^{-2}X^-X^3 \\
X^-X^+ &  = &  X^+X^- + \l X^3X^3 \nn
\eea
here $ \l = q-{1 \over q}$ .
It is natural to define a metric $g_{AB}$ and an invariant product $X\circ  Y$
\bea X\circ Y & = & g_{AB}X^AY^B \\ \nn
g_{+-}=-q ,\qquad   g_{-+} & = & -1/q ,\qquad   g_{33}=1
\eea
which let $X\circ X$ commute with $X^A$.
$P_3$ can be expressed through a generalized $\e$-tensor
\be {P_3}^{AB}{}_{CD}= {1 \over {1+q^4}}\e ^{FAB}\e_{FDC}
\ee 
where its indices are moved according to formulae like
\be \e_{ABC}=g_{CD}\,  \e _{AB}{}^D
\ee 
\bea \e_{+-}{}^3  =  q,\qquad  \e{-+}{}^3 &  = & -q,\qquad  \e_{33}{}^3  =
 1-q^2,   \nn \\
\e_{+3}{}^+  =  1, \qquad \e_{3+}{}^+ & = & -q^2, \\
\e_{-3}{}^-  =  -q^2, \qquad \e_{3-}{}^- & = & 1 \nn 
\eea 
Eq. (2.3) is then equivalent to 
\be X^CX^B\e_{BC}{}^A=0
\ee
and the R matrix can be expressed in the form
\be \hat {R}_{CD}^{AB}=\de_C^A\de_D^B-q^{-4}\e^{FAB}\e_{FDC}-q^{-4}(q^2-1)g^{AB}g_{CD}
\ee
Now we come to the definition of conjugation. We still choose
\be \overline{X^A}=g_{AB}X^B \equiv  X_A
\ee
like in paper [1] . But for generic complex $q$ this is consistent with
eqns. (2.3) only if we define $\bar{q}=q$ which means $q$ is unchanged under
conjugation. This forces us to distinguish between $q$ (and its functions) and
constant complex numbers which are to be conjugated as usual. (We mean
e. g. the $i$ in the Heisenberg relation, s. b.)

That is done best if the vector space the $q$-deformed operators act on is
considered as a (right) module over an algebra $A$ . This associative (and
commutative) algebra $A$ over the complex numbers is generated by the powers
of $q$ : $q$, $q^2$, ... $q^{r-1}$ and the condition $q^r=-1$ . The integer
$r$ is taken larger than 2  . Within $A$ we define an involution * which
fulfills the usual conditions 
\bea a^{\ast \ast}=a,   (ab)^{\ast}=b^{\ast} a^{\ast} \nn  \\
    (\a a+\b b)^{\ast}=\overline{\a}a^{\ast}+\overline{\b}b^{\ast}  
\eea 
where $\a$, $\b$ $\in$ $C$ .
Those properties are consistent with the choice $q^{\ast}=q$ which determines the
involution for all elements.

As a next step we consider a right module $M$ over the algebra $A$ . (Since
$A$ is commutative an equivalent approach is given considering a left module.)
$M$ is a complex vector space. For any $a, b \in A$ and $\h, \x \in M$ we have
\bea  \h (ab)&=&(\h a)b \nn \\
      \h(a+b)&=&\h a+\h b \\
      (\h+\x)&=&\h a+\x a \nn 
\eea
and any combin ation of type $\h a$ is again an element of $M$. For further
application we need a hermitean structure which is created by a hermitean
inner product. For any pair of elements a bilinear map $<\h\mid \x> \in A$  is
defined with the properties
\bea <\h\mid \x>^{\ast}&=&<\x \mid\h> \nn \\
     <\h a\mid\x b>&=&a^{\ast}<\h\mid\x>b
\eea
A third property, usually required for a hermitean product, includes the
absence of zero norm states. We shall see below that such states cannot be
excluded for our choice of $q$. Therefore, strictly speaking, our structure
is not hermitean in the usual sense. Nevertheless we keep this terminology but
remember that all unusual properties are connected with the existence of zero
norm states. The product allows the definition of the hermitean conjugation
$\overline{O}$  of an operator $O$ 
\be <\h\mid O\x>=<\overline{O}\h\mid\x>
\ee
We use another symbol not to mix this with the involution in $A$.
Subsequently hermitean and unitary (isometric) operators are
defined. Operators in $M$ can be viewed as matrices with entrances from $A$,
hermitean conjugation is then transposition together with the involution in
$A$ defined above. It is then clear that if $\l$ is an eigenvalue of $O$ then
$\l^{\ast}$ is an eigenvalue of $\overline{O}$ and hence $\l^{\ast} =\l$ for all
eigenvalues of an operator with $\overline{O}=O$ . We shall see below that for
our $q$ and the operators we are considering it is not necessary to
distinguish between hermitean and self-adjoint operators. Their eigenvalues
are real functions of $q$. One can show directly that the eigenvectors are
orthogonal (under the product defined above) for different eigenvalues with
the usual arguments. If a unitary operator has an eigenstate $\x$ with
eigenvalue $\l$ one can easily show 
\be <\x\mid\x>=\l^{\ast}\l<\x\mid\x>   
\ee
which gives information about $\l$ only for states with non-vanishing
norm. This fact becomes important below.

Based on eq (2.10) we can now proceed as in [1] and define a
derivative, momentum, angular momentum and the scaling operator $ \L$ in the same
way. For the components of the momentum we have the analog of (2.8), while for
angular momentum
\be L^CL^B\e_{BC}{}^A=-1/{q^2}WL^A
\ee
and \bea q^4(q^2-1)^2L\circ L & = & W^2-1 \nn \\
L^AW &  = & W L^A 
\eea
The scaling operator $\L$ is introduced in the same way with the properties
\bea \L^{1/2}X^A &  = & q^2X^A \L^{1/2} \nn \\
\L^{1/2}P^A & = & q^{-2}P^A\L^{1/2} \nn \\
\L^{1/2}L^A & = & L^A\L^{1/2} \\
\L^{1/2}W & = & W\L^{1/2} \nn
\eea
Conjugation of vector values is analogeous to eq. (2.10), $W$ is self-adjoint
and $\L$ is unitary up to normalization:
\be \overline{\L^{1/2}}=q^{-6}\L^{-1/2}
\ee  
Eqns. (2.16) lead to the standard $SO_q(3)$ algebra. The generalized
Heisenberg relations are 
\be P^A X^B-{\hat {R}^{-1}}{}^{AB}{}_{CD}X^CP^D=-{ i\over 2}
\L^{-1/2}\{(1+q^{-6})g^{AB}W-(1-q^{-4})\e^{ABC}L_C\}
\ee
Now we have to study representations of this algebra. For $q$ being a root of
unity the physical relevant representations become finite dimensional while
for real $q$ they have infinite dimension. Thus there is no difference here between
self-adjoint, essentially self-adjoint and hermitean operators.

The representations will be studied in detail in Chap. 4 .

\section{Representations of a one-dimensional $q$-deformed Heisenberg algebra}
We consider now a one-dimensional example of a $q$-deformed Heisenberg
algebra. That is neither a projection of the Euclidean space nor based on the
deformation of any symmetry group. It is even not non-commutative in the sense
of space coordinates because there is only one. Nevertheless it is based on a
modified Leibniz rule and has been studied for real $q$ in great detail
[5,6]. It reflects very nicely the deep role which is played by the scaling
operator $\L$ that one has to introduce in a general non-commutative structure
of coordinates and momenta. The algebra looks as follows :
\bea {1 \over \sqrt q}PX-\sqrt q XP & = & -iU \nn \\
UP & = & qPU \\
UX & = & {1 \over q}XU \nn
\eea
Conjugation is given by 
\bea \bar P & = & P \nn \\
\bar X & = & X \\
\bar U & = & U^{-1} \nn
\eea
While there is obviously no problem for real $q$, with our definition of
conjugation for operators and involution of algebra elements eq. (3.2) is also
consistent with (3.1). To give meaning to operators in our module space we
have to enlarge our algebra $A$ to include real functions of $q$ in a
straightforward way.                                   
We shall consider a representation of the algebra (3.1) based on eigenvectors
of $P$ . From the second equation it follows that applying $U$ to such an
eigenstate we obtain another one with eigenvalue multiplied by \\ $q^{-1}$ .
Therefore we have 
\be P\mid n>^{\p_0}= \p_0q^n\mid n>^{\p_0}
\ee
where n is integer, $0\leq  n\leq  2r-1$ , and $\p_0$ is an arbitrary real
function of $q$ .  Further
\be U\mid n>^{\p_0}= \mid n-1>^{\p_0}               
\ee 
and according to what was stated in last Chapter
\be \ ^{\p_0}<n\mid m>^{\p_0}=\de_{nm}
\ee
Now we have an example that the self-adjoint operator $P$ has eigenvalues
being real functions of $q$ .  The powers occuring are a 
consequence of the properties of $U$ . For our $q$ choosen we can see that the
eigenstate $U\mid 0>^{\p_0}$ has the same eigenvalue as $\mid 2r-1>^{\p_0}$.
Disregarding the case of degeneration we have 
\be U\mid 0>^{\p_0}=C(\p_0)\mid 2r-1>^{\p_0}
\ee
where $C$ is a phase factor and different $C$ label different
representations. From eqns. (3.4) and (3.6) we have $U^{2r}=C$ for any state
in our representation. Now it is straightforward to define another unitary
operator $U'$ by
\be U'=Ue^{-{i\a \over 2r}}
\ee
where we have put $C=e^{i\a}$
Then $U'^ {2r}=1$ and it is more convenient to work with a new system
$\mid n>'$
\be U' \mid n>' = \mid n-1>'
\ee
The new eigenstates are just multiplied by phase factors. For shortness we have
omitted the upper index $\p_0$ . From the first equation of (3.1) and its
conjugate one can deduce
\be XP=
{i \over \l}(\sqrt{ q}U-{1 \over \sqrt {q}} U^{-1}) 
\ee
which shows the action of $X$ on the states $\mid n>'$ states:
\be X\mid n>'={i \over q^n\l\p_0}(\sqrt{q}e^{i\a \over 2r}\mid n-1>'-{1 \over
  \sqrt{q}}e^{-{i\a \over 2r}}\mid n+1>')
\ee
 This system of $2r$ equations can be solved in principle and the eigenvalues und
 eigenstates of $X$ can be found. But it is easier to exploit the eigenstates
 of $U$, as we shall demonstrate below. We start with 
\bea \mid \f_0>& =& \sum _{n=0}^{2r-1}\mid      n>' \\
\mid \f_k>& =& (\p_0)^{-k}P^k\mid \f_0>=\sum _{n=0}^{2r-1} q^{kn}\mid n>' \nn
\eea
and integer $0\leq k \leq 2r-1$ . Obviously 
\be U'\mid \f_k>=q^k\mid \f_k>
\ee
We mention that for real $q$ those states are non-normalizable which is not
the case here.

Before constructing the eigenstates of $X$ we shortly comment on the
eigenstates of $U'$ and $U$ . Our definition of an adjoint operator in Chap. 2
and the inner product lead to unitary operators with
respect to that product which will have properties differing from the usual
ones, as we have already seen for self-adjoint operators. The eigenstates of
our unitary operators may not be orthogonal and can contain zero norm
states. So explicitly
\be <\f_k\mid \f_m>=\sum _{n=0}^{2r-1} q^{n(k+m)}
\ee
what is non-zero for $m=k=0$ or $m+k=2r$ . Hence we have only two non-zero
norm states for $k=0$ and $r$ and the eigenstates $\mid \f_k>$ and $\mid
\f_{2r-k}>$ for $k=1 , \ldots , r-1$ are not orthogonal.
 
 Keeping in mind all that we can
still work with the states (3.11) as a basis to construct the $X$ eigenstates.

From the algebra (3.1) follows
\be X\mid \f_k>=d_k\mid \f_{k-1}>
\ee
for $1\leq k \leq 2r-1$ and
\be X\mid \f_0>=d_0\mid     \f_{2r-1}>
\ee
Next we have to calculate $d_k$ . We apply the conjugate of eq. (3.9) to $\mid
\f_k>$ and find
\be d_k={i \over {\l \p_0}}(e^{i\a \over {2r}}q^{k-{1\over 2}}-e^{-{i\a \over
    2r}}q^{-k+{1\over 2}})
\ee
This formula works for all $0\leq k\leq 2r-1$ . We construct the eigenstates
the following way
\bea  X\mid x_m> & = & x_m\mid x_m> \nn \\
\mid x_m> & = & \sum  _{k=0}^{2r-1} a_k\mid \f_k>
\eea
yielding the recursion relation for the coefficients
\be a_{k+1}={x_m \over d_{k+1}}a_k
\ee
Consistency requires
\be a_0={x_m \over d_0}a_{2r-1}
\ee
We can put $a_0=1$ and the solution of eqs. (3.18) and (3.19) are
\bea a_k& =& (x_m)^k ( \prod _{l=1}^{k}d_l)^{-1} \nn \\
(x_m)^{2r}& =& \prod _{l=1}^{2r}d_l={i^{2r}\over \l^{2r}{\p_0}^{2r}}(-1)^rf^2(q,\a)
\eea
where we have introduced the function
\be f(q,\a)=\prod _{k=1}^{r}(q^{k+{1\over 2}}e^{i\a\over {2r}}-q^{-k-{1\over
  2}}e^{-{i\a\over {2r}}})
\ee
Eq. (3.20) gives (in principle) the possibility to find the eigenvalues of
$X$. They depend on $\p_0$ and the real function $f^2(q,\a)$ . The fact, that
only $(x_m)^{2r}$ is given, reflects the property that due to the unitary
equivalence of $X$ and $P$ $x_m$ must be proportional to $q^m$ . Thus
eq. (3.20) contains no new information we did not have before. The function
$f(q,\a)$ occurs also in the more realistic three-dimensional case (s. next
Chapter) .  

Now we can compare our results with those for real $q$ obtained in papers [5]
and [6]. The main difference is that all our representations have finite
dimensions which avoids the mathematical problems of the real case.  On the other hand
we have to introduce an additional parameter $C$ (or $\a$) characterizing the
representation. The operators $X$ and $P$ are manifestly
equivalent in our representation. 

\section{$SO_q(3)$ deformation in compact space}
In this Chapter we give the representations of the $q$-deformed algebra (2.8),
(2.16) - (2.20) for $q^r=-1$ . We have not written the $L^AX^B$ and
$L^AP^B$ relations which are the same as in [1] . We are also not going to
repeat the derivations of papers [1] and [7] leading to the $T$-operators and
explaining the appearance of the Clebsch-Gordon coefficients because on the
algebraic level there are no changes. The changes start as soon as
representations are considered, what shall be done now.

We choose $L\circ  L$, $ L^3$ and $X \circ X$  as a complete set of commuting
variables. One can proceed as in the undeformed case and exploit eqs.
(2.16) and (2.17) . For the angular momentum the eigenvalues are
\bea L\circ L\mid j,m,n>={q^{-6}\over
  (q^2-q^{-2})^2}(q^{4j+2}+q^{-4j-2}-q^2-q^{-2})\mid j,m,n> \nn \\
L^3\mid j,m,n>=-{q^{-3}\over (q-q^{-1})}(q^{2m}-{q^{2j+1}+q^{-2j-1}\over
  q+q^{-1}})\mid j,m,n>
\eea
where $j$ and $m$ are integers, $|m|\leq j$ and $0\leq j\leq j_{max}$
. (Note that the sign of $L^3$ is opposite to the usual one , because we have
kept the conventions of paper [1].) 
For $q$ being a root of unity we must remember that there are two types of
representations, called types I and II in paper [3]. We allow only type II
representations for the construction of the non-commutative space. That the
type I representations can be omitted consistently follows from paper [4]. The
type II representations behave as for $q=1$ (and general real $q$) except the
fact $j_{max}\leq {{r\over 2}-1}$ . The states are fully determined by the
quantum numbers  $j$, $m$ and $n$ . From the first eq. of (2.13) we read off
\be X^2\mid  j,m,n>=l^2_0q^{4n}\mid j,m,n>
\ee
It is sufficient to choose the integer $n$ as  $0\leq n\leq {r-1}$  . The
parameter $l_0$    plays the same role as $\p_0$ in the one-dimensional case.

All our representations are unitary and either irreducible or fully reducible
[3]. Irreducible representations are labelled by the integer $j$ . Because of
eq. (4.2) we deal with  finite dimensional irreducible representations like in
the one-dimensional case before. That and the existence of a $j_{max}$ are the
main differences with respect to real $q$ .

The states are normalized in the usual way. The
phase factors can be choosen to fulfill
\bea \L^{1\over 2}\mid j,m,n>& =& q^{-3}\mid j,m,n-1> \nn \\
\L^{-{1\over 2}}\mid j,m,n>& =& q^3\mid j,m,n+1>
\eea
From eq. (2.16) the matrix elements of $L^\pm  $ can be obtained. We mention for
further use
\be W\mid j,m,n>\ ={\{2j+1\}\over \{1\}}\mid j,m,n>
\ee
where we have introduced the abbreviations
\bea \{a\} & = & q^a+q^{-a} \nn \\
\left[  \ a \ \right] & = & {q^a-q^{-a}\over \l}
\eea
In papers [1] and [7] one finds how the $SO_q(3)$ structure can be used to
define reduced matrix elements for $X^A$ and $P^A$ . For the non-vanishing
matrix elements we quote the results 
\bea <j+1,m+1,n\mid X^+\mid j,m,n>=q^{m-2j}\sqrt{\left[ j+m+1\right] \left
    [ j+m+2\right] }<j+1,n\parallel X^-\parallel j,n> \nn \\   
<j-1,m+1,n\mid X^+\mid j,m,n>=q^{m+2j+2}\sqrt{\left[ j-m\right] \left
    [ j-m-1\right] }<j-1,n\parallel X^-\parallel j,n> \nn \\
<j+1,m-1,n\mid X^-\mid j,m,n>=q^m\sqrt{\left [j-m+1\right] 
   \left[ j-m+2\right] } <j+1,n\parallel X^-\parallel j,n> \nn \\
<j-1,m-1,n\mid X^-\mid j,m,n>=q^m\sqrt{\left[ j+m\right] \left[ j+m-1\right]
}<j-1,n\parallel X^-\parallel j,n> \nn \\
<j+1,m,n\mid X^3\mid j,m,n>=q^{m-j-{1\over
    2}}\sqrt{1+q^2}\sqrt{\left[j-m+1\right]    \left[ j+m+1\right]
  }<j+1,n\parallel X^-\parallel j,n> \nn \\
<j-1,m,n\mid X^3\mid j,m,n>=-q^{m+j+{1\over 2}}\sqrt{1+q^2}\sqrt{\left
      [ j-m\right] \left[   j+m\right] }<j-1,n\parallel X^-\parallel j,n>
\nn \\
\eea
The matrix elements on the r.h.s. are the reduced ones. Using conjugation
properties (2.10) we have
\be <j+1,n\parallel X^-\parallel j,n>=-q^{2j+2}\overline{<j,n\parallel
  X^-\parallel j+1,n>}
\ee
Therefore only one reduced matrix element has to be determined what is easily
obtained from the first eq. of (2.3) and (4.2). We fix the phase by setting
\be <j+1,n\parallel X^-\parallel j,n>={l_0q^{j+2n}\over \sqrt{\left[ 2\right]
    \left[ 2j+1\right] \left[ 2j+3\right] }}
\ee
By the way, the first eq. of (2.3) also tells us that $<j,n\parallel
X^-\parallel j,n>$ must vanish.

Now we come to the matrix elements of $P^A$ . Based on eqs. (4.6) and (4.7)
they are calculable relying on the matrix elements of the values $X\circ P$
and its conjugate $P\circ X$ . The Heisenberg relation (2.20) with the help of
the $R$ matrix (2.9) yields after contraction
\be P\circ X -q^6X\circ P = -{i\over 2}\l^{-{1\over
    2}}(1+q^{-6})(q^2+1+q^{-2})W
\ee
Together with its conjugation eq. (4.9) gives
\bea X\circ P&=&-{i\over 2} {(\L^{1\over 2}-\L^{-{1\over 2}})W \over q^2(q^2-1)}
  \nn \\
P\circ X&=&{i\over 2} {(q^{-6}\L^{-{1\over 2}}-q^6\L^{1\over 2})W \over q^2(q^2-1)}
\eea
Therefore $X\circ P$ has matrix elements only between neighbouring $n$. We
consider now
\bea <j,m,n\mid X\circ P\mid j,m,n+1>=&-&q^2\{ \left[ 2j+3\right] \left[
    2j+2\right] \nn \\ 
 &<&j,n\parallel X^-\parallel   j+1,n> <j+1,n\parallel P^-\parallel j,n+1> \nn  \\
&+& \left[ 2j\right]  \left[ 2j-1\right] \nn \\ & <&j,n\parallel X^-\parallel
j-1,n><j-1,n\parallel P^-\parallel j,n+1>\} \nn \\
=&-&{i\over 2}{W_j\over q^5(q^2-1)} 
\eea
where the reduced matrix elements of $P^A$ are defined analogeous to
eqs. (4.6) including the fact that they are no longer diagonal in $n$. Now it
is straightforward to take 
\bea <j,m,n+1\mid X\circ P\mid j,m,n>=&-&q^2\{\left[ 2j+3\right] \left
  [ 2j+2\right] \nn \\  &<&j,n+1\parallel X^-\parallel j+1,n+1><j+1,n+1\parallel
P^-\parallel j,n> \nn \\
&+& \left[ 2j\right] \left[ 2j-1\right] \nn \\ &<&j,n+1\parallel X^-\parallel
j-1,n+1><j-1,n+1\parallel P^-\parallel j,n>\} \nn \\
=&{i\over 2}&{W_jq\over q^2-1} 
\eea
We put in eqs. (4.7) and (4.6) and the conjugation relations
\bea <j+1,n\parallel P^- \parallel j,n+1>&=&-q^{2j+2}<j,n+1\parallel P^-\parallel
j,n> \nn \\
<j+1,n+1\parallel P^-\parallel j,n>&=&-q^{2j+2}<j,n\parallel P^-\parallel j+1,n+1>
\eea
The system (4.11) and (4.12) can be rewritten as two recursion relations in
$j$ for the two unknowns, the reduced matrix elements of $P$ . An easy way to
solve it , is to start with $j=0$ , read off the general formula and prove it
by insertion. For clearness, we present all non-vanishing reduced matrix
elements
\bea <j+1,n\parallel P^-\parallel j,n+1>&=&-iq^{-j-6-2n}Z^{-1}, <j,n+1\parallel
P^-\parallel j+1,n>=-iq^{-3j-8-2n}Z^{-1} \nn \\
<j+1,n+1\parallel P^-\parallel j,n>&=&iq^{3j-2-2n}Z^{-1}, <j,n\parallel
P^-\parallel j+1,n+1>=iq^{j-4-2n}Z^{-1}
\eea 
where the common denominator is
\[ Z=2l_0\l \sqrt{\left[ 2\right] \left[ 2j+1\right] \left[ 2j+3\right] } \]
Neither eq. (4.8) nor eq. (4.14) contains any divergencies because of the
condition $j_{max}\leq {r\over 2}-1$ . If $j+1$ exceeds $j_{max}$ the matrix
element simply vanishes as it does for $j-1=-1$.

Our next aim is to calculate the eigenvalues of $P^2\equiv P\circ P$ . We shall
follow the lines of Chap. 3 and start with the definition of a unitary
operator
\be U=q^3\L^{1\over 2} \ee
Going back to eq. (4.3) we have
\be U\mid n>=\mid n-1> \ee
where we have omitted all quantum numbers which are unchanged.
After 
\be U\mid 0>=e^{i\a}\mid r-1> \ee
we introduce
\bea U'&=&Ue^{-{i\a\over r}} \nn \\
U'\mid n>'&=&\mid n-1>'
\eea
The eigenstates of the operator $U'$ are given by
\bea \mid \f_k>&=&\sum _{n=0}^{r-1} q^{2nk}\mid n>' \nn \\
U'\mid \f_k>&=&q^{2k}\mid \f_k>
\eea
Note that the eigenstates for even $k$ can be produced by the operator $X^2 /
{l_0}^2$ acting $k/ 2$ times on $\mid \f_0>$ . From the algebra (2.18) follows
\be P\circ P\mid \f_k>=\tilde {d_k}\mid \f_{k-2}> \ee
where we shall calculate $\tilde {d_k}$ below. For the $P$-eigenstates we use
the ansatz
\bea P\circ P\mid p_n>&=&p^2_n\mid p_n> \nn \\
\mid p_n>&=&\sum _{k=0}^{r-1} a_k\mid \f_k>
\eea
Eq. (4.20) yields the recursion relation
\be a_{k+2}={p^2_m\over \tilde {d}_{k+2}}a_k \ee
Now it is necessary to distinguish between even and odd $r$ . In the first
case we obtain two different solutions putting $a_0=1, a_1=0$ and vice
versa. They contain either even or odd numbers of $k$ in the sum
(4.21). Consistency gives for the eigenvalues
\bea (p^2_+)^{r\over 2}&=&\prod _{k=0}^{{r\over 2}-1} \tilde {d}_{2k} \nn \\
(p^2_-)^{r\over 2}&=&\prod _{k=0}^{{r\over 2}-1} \tilde {d}_{2k+1} 
\eea
For odd $r$ the sum (4.21) contains all numbers and hence 
\be (p^2)^r=\prod _{k=0}^{r-1} \tilde {d}_k \ee
The coefficients $\tilde {d}_k$ are calculated via the matrix elements of
$P^2$ between the $\mid j,n>$ states. We have the same structure as in the
first parts of eqs. (4.11) and (4.12), e.g.
\bea <j,n+2\mid P^2\mid j,n>=&-&q^2\{ \left[ 2j+3\right] \left[ 2j+2 \right] \nn
\\ & <&j,n+2\parallel P^-\parallel j+1,n+1><j+1,n+1\parallel P^-\parallel
j,n>\nn \\ & +&\left[ 2j\right[ \left[ 2j-1\right]  \\
&<&j,n+2\parallel P^-\parallel j-1,n+1><j-1,n+1\parallel P^-\parallel j,n>\}\nn 
\eea 
With the results of eq. (4.14) we get 
\be <j,n+2\mid P^2\mid j,n>=-{q^{-4n-10}\over 4l^2_0\l^2} \ee
and the same way
\be <j,n-2\mid P^2\mid j,n>=-{q^{-4n-2}\over 4l^2_0\l^2} \ee
A little bit more lengthy is the calculation of the diagonal element due to
the doubling of terms connected with intermediate states having quantum
numbers $n\pm 1$ .
\be <j,n\mid P^2\mid j,n>={q^{-4n-6}\over 4l^2_0\l^2} \{4j+2\} \ee
As soon as the quantum numbers of the r.h.s. ket vector are fixed there are no
further non-vanishing matrix elements. Now we consider
\bea P^2\mid n>'&=&\mid n>' <n\mid ' P^2\mid n>' \nn \\ & +& \mid n+2>'
<n+2\mid ' P^2\mid
n>' + \mid n-2>' <n-2\mid ' P^2 \mid n>'  \\
&=&\mid n>' <n\mid P^2\mid n> \nn \\ & +& \mid n+2>' e^{-{2i\a\over r}}<n+2\mid  P^2 \mid  
n> + \mid n-2>' e^{2ia\over r}<n-2\mid P^2\mid n> \nn
\eea
From eq. (4.20) follows
\bea P^2\mid \f_k>&=&\sum_{n=0}^{r-1} q^{2nk}\mid n>'(q^{-4k}e^{-{2i\a\over
    r}}<n+2\mid  P^2\mid n>+q^{4k}e^{2i\a\over r}<n-2\mid P^2\mid n > \nn \\
&\,&+<n\mid P^2\mid n>) \nn \\
&=&\tilde {d}_k\sum_{n=0}^{r-1} q^{2nk-4n}\mid n>' 
\eea
Substituting eqs. (4.26)-(4.28) we can read off
\bea \tilde {d}_k & = & {q^{-6}\over 4l^2_0\l^2}(\{ 4j+2\} -q^
{ -4k-4}e^{-{2i\a\over r}} - q^{4k+4}e^{2i\a\over r})  \nn \\
 & = &- {q^{-6}\over 4l^2_0}\left[ 2k+2j+3\right]_{\a} \left[
   2k-2j+1\right]_{\a}  
\eea
where we have introduced the abbreviation
\be \left[a\right]_{\a}={{q^a e^{i\a\over r}-q^{-a} e^{-{i\a\over r}}}\over \l}
\ee
Finally we have for
even $r$
\be (p^2_\pm)^{r\over 2}={-i^r\over 2^r\l^rl^r_0}(-1)^{r\over
  2}f(q,\a)f(q,\a-\p r)
 \ee
and for odd $r$
\be (p^2)^r={-i^{2r}\over 2^{2r}\l^{2r}l^{2r}_0}f^2(q,\a)f^2(q,\a-\p r)
 \ee

While $\tilde {d}_k$ depends on $j$ \, $p^2$ , of course, does not. It is
remarkable that eqs. (4.33) and (4.34) very much resemble eq. (3.20) derived for the
one-dimensional model.

For even $r$ any eigenvalue is degenerated twice, disregarding the obvious
degeneration with respect to $j$ and $m$ . All
eigenvectors (4.21) are orthogonal and normalizable. (Note that this is not
true for the $\mid \f_k>$  states.) The eigenvalues of $P^2$ are in both cases
given by even powers of $q$ multiplied by the roots of real functions on $q$ .
The main difference to real $q$ is the finiteness in dimension for the
eigenvector space.

It would be interesting to know more about the function $f(q,\a)$ esp. it
should play a role in a generalized Fourier transformation. We hope to return
to this question in our further work. Our experience for finite $r$ seems to
lead to the conjecture that the roots in eqs. (4.33), (4.34) can be easily
extracted if one excludes all polynomials which vanish after being multiplied
with non-zero combinations of powers of $q$ .

At the end of this section we shall return to the question, how the structure
we have found is related to former attempts of combining non-commutative
geometry with string theory via a matrix realization [8].

Our operators $X^i$ can be viewed as matrices acting on vectors with dimension
${r^3}\over 4$ (for even $r$) as long as $l_0$ is kept fixed. It is natural to
ask whether they can be considered as analogues of the $X_i$ fields ($ 0\leq
i\leq 9$) in the IKKT model [8]. The role played there by $SO(10,C)$ is here
played by $SO(3)$ .

Nevertheless there are substantial differences between the two sets of
operators. Even though there is an analogue of their unitary gauge fields
$U_i$ namely  the unitary operator $q^3\L^{1\over 2}$ (it is only one), that
operator plays a different role. Its "gauge transformation" induces a
multiplicative rescaling while in paper [8] the gauge transformation adds a
constant (proportional to the compactification radius). It is that fact which
requires infinite matrices in the IKKT model in order to have an infinite
trace, while direct calculation yields $Tr X^i=0$ in our case. This
discrepancy becomes less important remembering that for the full trace one has
to integrate over $l_0$ which produces a divergent result.

We propose to solve the remaining problems by taking into account the fact,
that in the IKKT model the undeformed group $SO(10,C)$ was used while we
started with the $q$-deformed $SO_q(3)$. 

Concluding this remark we state, that we have found a self-consistent
structure which is close to become an analogue of some IKKT like matrix model
on a non-commutative torus. This problem is under work now.

\section{Acknowledgements}
This work has been supported by Alfried Krupp von Bohlen und
Halbach--Stiftung, \\ Essen . \\
The author thanks J. Wess, D. L\"ust and M. Karowski for helpful discussions.

\end{document}